\documentclass[%
reprint,
showpacs,preprintnumbers,
 amsmath,amssymb,
 aps,
pra,
]{revtex4-1}

\usepackage{graphicx}
\usepackage{dcolumn}
\usepackage{subfigure}
\usepackage{bm}
\usepackage{hyperref}
\usepackage{color,soul}
\usepackage{xcolor}	

\usepackage{bbold}

\newcommand{\ket}[1]{|#1\rangle}
 \newcommand{\bra}[1]{\langle #1|}

 \newcommand{\T}{{\mathrm t}}

\newcommand{\Tr}{{\mathrm {Tr}}}
\newcommand{\diag}{{\mathrm {diag}}}
\newcommand{\spa}{{\mathrm {span}}}

\newcommand{\etal}{\textit {et al. }}

\begin{document}

\renewcommand*{\bibname}{Refrences}


\title{ Parametrization of quantum states and  the  quantum state discrimination problem  }

\author{Seyed Arash Ghoreishi}
\affiliation{Department of Physics, Ferdowsi University of Mashhad, Mashhad, Iran}
\author{Seyed Javad Akhtarshenas}
 \email{akhtarshenas@um.ac.ir}
\affiliation{Department of Physics, Ferdowsi University of Mashhad, Mashhad, Iran}
\author{Mohsen Sarbishaei}
\affiliation{Department of Physics, Ferdowsi University of Mashhad, Mashhad, Iran}

\begin{abstract}
A discrimination problem consists  of $N$ linearly independent  pure quantum states $\Phi=\{\ket{\phi_i}\}$ and  the corresponding  occurrence probabilities $\eta=\{\eta_i\}$. To any such problem   we associate, up to a permutation over the probabilities $\{\eta_i\}$,  a unique pair of  density matrices  $\boldsymbol{\rho_{_{T}}}$ and $\boldsymbol{\eta_{{p}}}$ defined on the $N$-dimensional Hilbert space $\mathcal{H}_N$. The first one, $\boldsymbol{\rho_{_{T}}}$,   provides a new parametrization of a generic full-rank density matrix in terms of the parameters of the discrimination problem, i.e. the mutual overlaps $\gamma_{ij}=\bra{\phi_i}\phi_j\rangle$ and  the occurrence probabilities $\{\eta_i\}$. The second one on the other hand is  defined as a  diagonal density matrix $\boldsymbol{\eta_p}$ with the diagonal entries given by the probabilities $\{\eta_i\}$ with the ordering induced by the permutation $p$ of the probabilities.  $\boldsymbol{\rho_{_{T}}}$ and $\boldsymbol{\eta_{{p}}}$   capture  information about the quantum and classical versions of the discrimination problem, respectively. In this sense,  when the set $\Phi$  can be discriminated unambiguously with probability one, i.e. when the states to be discriminated are mutually orthogonal and can be distinguished by a classical observer,   then $\boldsymbol{\rho_{_{T}}} \rightarrow \boldsymbol{\eta_{{p}}}$. Moreover, if the set lacks its independency and  cannot be discriminated anymore  the distinguishability of the pair, measured by the fidelity $F(\boldsymbol{\rho_{_{T}}}, \boldsymbol{\eta_{{p}}})$,  becomes minimum.  This enables one to associate to each discrimination problem  a measure of discriminability defined by the fidelity $F(\boldsymbol{\rho_{_{T}}}, \boldsymbol{\eta_{{p}}})$.  This quantity, though distinct from the  the maximum probability of success, has the advantage of being easy to calculate  and in this respect it can find useful applications  in estimating the extent to which  the set is discriminable.

\end{abstract}



\maketitle


\section{Introduction}\label{s1}
Due to the increasing interest in the emerging field of quantum information theory, much  attention has been paid to the investigation of the set of quantum states,  both from theoretical and practical point of view. From a theoretical perspective, it is important to find a method to describe and characterize  the set of $N$-level quantum states, and   a lot of work has been devoted for such purposes \cite{FanoRMP1957,BoyaARXIV1998,ByrdPLA2001,TilmaJPA2002,TilmaJPA2002-2,TilmaJGP2004,
ZyczkowskiJPA2001,DitaJPA2005,DitaJPA2003,AkhtarshenasJMP2007,AkhtarshenasJPA2007,IlinJPA2018}.
Apart from theoretical investigations of the set of mixed states,   pure quantum states    are also the focus of attention for their particular role in  information theory.    They provide the key  ingredient to encode quantum information and in this respect can be considered as the resource of  quantum protocols.   The existence of nonorthogonal states, arising from superposition, is crucial in quantum phenomena  makes quantum information  so different from the classical information.  It follows that there must exist states  with nonzero overlap that cannot be discriminated perfectly.  The  discrimination of quantum states is of particular importance in  quantum information theory, especially  in quantum communication \cite{HolevoBook1982,HelstromBook1976} and quantum cryptography \cite{GisinRMP2002}.    This property  of nonorthogonal states  could also be responsible for some nonclassicalities  such as quantum discord \cite{ZurekPRL2001,HendersonJPA2001}, no-cloning theorem \cite{WootersNature1982,BuzekPRA1996} and  no-local-broadcasting \cite{PianiPRL2008}.

A general   quantum state is represented by a Hermitian positive semi-definite matrix $\rho$ with the unit trace and can be  realized  as  a  statistical ensemble of pure states. By statistical ensemble we mean a  set of pure states with a certain  probabilities.  Although with a given  statistical ensemble it is easy to  write its associated  density matrix  by means of convex combination, the converse is not true, i.e. having a density matrix it makes no sense to ask  which ensemble the state is constructed from. This is as a result of the unitary freedom in the ensemble for density matrices \cite{NielsenBook2010}, i.e.   an infinite number of different ensembles can have the same density matrix,  in turn, implies  that a density matrix corresponding to a given ensemble does not tell about  what ensemble it uniquely  corresponds to.  Now the question arises that; can we associate a kind of density matrix  with  a set of pure states and corresponding prior probabilities in such a way that it captures all the information about the set? Such a density matrix, if possible to construct, will be different from the usual one constructed by means of convex combination, meaning that  it does not exploit  the corresponding set of states as a realization ensemble. Here we address this question and  show that one can always define such a kind of density matrix provided that the set of states forms a linearly independent set of pure states with a priori probability of occurrence. Our result is tied, on the one side,  to the discrimination problem of a set of states and, on the other side, it provides a way to parameterize a general density matrix.  We therefore introduce  a new parametrization of density matrices using the problem of quantum state discrimination.

There are two schemes for quantum state discrimination; minimum-error discrimination and unambiguous state discrimination \cite{BarnettAOP2009}. In the minimum-error scheme there is always an answer with a non-zero probability of wrong detection. In this strategy the task is to minimize the probability of occurring error \cite{HelstromBook1976,BarnettJPA2009}. Some examples for which explicit results are known include symmetric states with equal a priori probability \cite{BanIJTP1997}, three mirror symmetric qubit states \cite{AnderssonPRA2002,ChouPRA2004}  and discrimination between subsets of linearly dependent quantum states \cite{HerzogPRA2002}.
In the unambiguous discrimination of quantum states \cite{IvanovicPLA1987,PeresPLA1988,DieksPLA1988}, on the other hand,  if the output $i$ corresponding to the state $\rho_i$ is detected by a receiver, one can claim with certainty  that the state is $\rho_i$; however, for nonorthogonal states, there is always a non-zero probability for failure in the detection of the state and the aim is to minimize this failure probability.  In 1998, Chefles has shown that a set of pure states can be unambiguously discriminated if and only if they are linearly independent \cite{CheflesPLA1998}.

In this paper, we follow the problem of quantum state discrimination and introduce a new parametrization for an arbitrary density matrix of $N$-level quantum system.  For this purpose, we consider a  set of linearly independent but not necessarily orthogonal  states  $\{\ket{\phi_i}\}$, with the prior probabilities $\{\eta_i\}$, and construct a linear invertible transformation $T^e_{\eta_p}$ that transforms  the original nonorthogonal set to the orthogonal set $\{\sqrt{\eta_i}\ket{e_i}\}$.  The elements of this  transformation matrix are expressed in terms of  the mutual overlaps $\gamma_{ij}=\bra{\phi_i}\phi_j\rangle$ and  the  occurrence probabilities $\eta_i$.
We then show that although  construction of such transformation matrix is not unique, it provides, up to a permutation of the probabilities $\eta_i$,  a unique pair of density matrices $\boldsymbol{\rho_{_{T}}}$ and $\boldsymbol{\eta_{{p}}}$. The first one, $\boldsymbol{\rho_{_{T}}}$,   provides a representation of a generic full-rank density matrix in terms of the parameters of the discrimination problem, i.e. the mutual overlaps $\gamma_{ij}=\bra{\phi_i}\phi_j\rangle$ and  the occurrence probabilities $\{\eta_i\}$. The second  one, on the other hand, is  defined as a  diagonal density matrix $\boldsymbol{\eta_p}$ with the diagonal entries given by the probabilities $\{\eta_i\}$ with the ordering induced by the permutation $p$ of the probabilities. This  pair  enables one to associate to each  problem of discrimination of a set of $N$ pure states the  corresponding problem of distinguishability of two density matrices defined on the $N$-dimensional Hilbert space $\mathcal{H}_N$.  By using the notion of fidelity as a measure of distinguishability of the two density matrices $\boldsymbol{\rho_{_{T}}}$ and $\boldsymbol{\eta_{{p}}}$, we define the notion of  discriminability  of the set of $N$ pure states $\{\ket{\phi_i}\}$ with the prior probability $\{\eta_i\}$.

The rest of the paper is organized as  follows. In Sec. \ref{section-2}, we construct the linear invertible operator $T^e_{\eta_p}$ and introduce a new parametrization for a general density matrix. In Sec. \ref{section-3}, we apply our results to introduce the notion of discriminability  and investigate its properties.  The paper is concluded in section IV with a brief discussion.

\section{Parameterizing quantum states}\label{section-2}
Consider the set $\Phi_{\eta}$  consists of a set   of linearly independent and normalized but not necessary orthogonal states $\Phi=\{\ket{\phi_i}\}_{i=1}^N$  with a priori probability $\{\eta_i\}_{i=1}^N$.  For the set $\Phi$ one can define the dual (reciprocal) set $\tilde{\Phi}=\{\ket{\tilde{\phi}_i}\}_{i=1}^N$ in such a way that \cite{HornBook2013}
\begin{equation}\label{Dual}
\bra{\tilde{\phi}_i} \phi_j\rangle =\delta_{ij}.
\end{equation}
Evidently, both sets span the same subspace  $V$, i.e. $V=\spa{\{\ket{\phi_i}\}}=\spa{\{\ket{\tilde{\phi_i}}\}}$.
 Let also  $\{\ket{e_i}\}_{i=1}^N$ serves an orthonormal basis for $V$.  For such orthonormal basis $\{\ket{e_i}\}_{i=1}^N$ and the set  $\Phi_{\eta}$, one can define a linear invertible transformation $T^e_{\eta_p}$  such that
\begin{equation}\label{T1}
T^e_{\eta_p}\ket{\phi_i}=\sqrt{\eta_{p_i}}\ket{e_i},
\end{equation}
where $p=\{p_1, \cdots,p_N\}$ is any permutation of $\{1,\cdots,N\}$. Indeed, for an orthonormal basis $\{\ket{e_i}\}_{i=1}^N$ and linearly independent vectors $\{\ket{\phi_i}\}_{i=1}^N$ with associated probability $\{\eta_i\}_{i=1}^N$, the linear transformation $T^e_{\eta_p}$ is unique up to a permutation over the set $\{\eta_i\}_{i=1}^N$.
More precisely
\begin{eqnarray}
T^e_{\eta_p}=\sum_{i}^N\sqrt{\eta_{p_i}}\ket{e_i}\bra{\tilde{\phi}_i},\quad  {T^{e^{-1}}_{\eta_p}}=\sum_{i}^N\frac{1}{\sqrt{\eta_{p_i}}}\ket{\phi_i}\bra{e_i}.
\end{eqnarray}

\subsection{Quantum state parametrization}
The linear operator $T^e_{\eta_p}$ transforms the linearly independent set of states $\Phi=\{\ket{\phi_i}\}_{i=1}^{N}$ to an orthonormal basis $\{\ket{e_i}\}_{i=1}^{N}$. Using the standard  Gram-Schmidt procedure, one can construct this particular basis in such a way that $T^e_{\eta_p}$ takes an upper triangular matrix representation \cite{ArfkenBook2013}. However, $\{\ket{e_i}\}_{i=1}^{N}$ does not provide an arbitrary basis obtained from $\Phi$.  Indeed, starting from a given linearly independent set $\Phi$ and for an arbitrary unitary operator $U$,  we can obtain an arbitrary orthonormal basis $\{\ket{u_i}=U\ket{e_i}\}_{i=1}^{N}$  by means of the invertible transformation  $T^u_{\eta_p}=UT^e_{\eta_p}$, i.e.
\begin{equation}\label{T2}
T^u_{\eta_p}\ket{\phi_i}=\sqrt{\eta_{p_i}}\ket{u_i}.
\end{equation}
Motivated by this, we define the projection map $\Pi$ and associate to  each linear transformation $T$ a density matrix $\boldsymbol{\rho_{_T}}$ as
\begin{equation}\label{PiMap}
\Pi \; : \quad \frac{T}{\sqrt{\Tr[T^\dagger T]}} \; \longrightarrow \; \boldsymbol{\rho_{_T}}=\frac{T^\dagger T}{\Tr[T^\dagger T]}.
\end{equation}
Clearly, the above map results in the same state $\boldsymbol{\rho_{_T}}$ if we change $T\longrightarrow T^\prime= UT$ for an arbitrary unitary transformation $U$.

To fix our terminology and giving a geometrical interpretation of the above discussion,   suppose $\mathcal{F}$ represents the set of all linear transformations $T^u_{{\eta_p}}$ that construct a general orthonormal basis $\{\ket{u_i}\}$ from an arbitrary  discrimination problem $\Phi_{\eta}$, following the route of Eq. \eqref{T2}.   Suppose  also that $\mathcal{M}$ consists of  all density matrices  $\boldsymbol{\rho_{_{T}}}$ that can be obtained by applying the map $\Pi$ on all transformations $T^u_{{\eta_p}}\in \mathcal{F}$. For the sake of simplicity we drop all the indices and use the simplified notation $\boldsymbol{\rho_{_{T}}}$ for the states associated to the transformations $T^u_{{\eta_p}}$, but it is clear that  $\boldsymbol{\rho_{_T}}$ depends both on the parameters of the discrimination problem and also on the particular permutation $p$. Since by definition $\mathcal{F}$ consists of  invertible transformations,  the elements of  $\mathcal{M}$ are  full rank density matrices defined on the $N$-dimensional Hilbert space $\mathcal{H}_N$.  With this terminology, the map \eqref{PiMap} can be regarded as the bundle projection map,  $\Pi :  \mathcal{F}\rightarrow \mathcal{M}$, with $\mathcal{M}$ as the base space and $\mathcal{F}$ as the bundle space.  This enables one to construct a fiber bundle structure such that  the set of all transformations $T^u_{{\eta_p}}$ that project to the same state $\boldsymbol{\rho_{_{T}}}$ is considered  as the fiber over $\boldsymbol{\rho_{_{T}}}$.

Looking at Eqs. \eqref{T1} and \eqref{T2},  one can see that if the original set $\Phi$ be orthonormal, i.e. $\ket{\phi_i}=\ket{e_i}$ for $i=1,\cdots,N$, then  the upper triangular matrix $T^e_{\eta_p}$ becomes a diagonal matrix
\begin{equation}
T^e_{\eta_p}\rightarrow \sqrt{\boldsymbol{\eta_p}}=\sum_{i=1}^N\sqrt{\eta_{p_i}}\ket{e_i}\bra{e_i},
\end{equation}
where $\boldsymbol{\eta_p}=\diag\{\eta_{p_1},\cdots,\eta_{p_n}\}$ is a diagonal density matrix  in the orthonormal basis $\{\ket{e_i}\}$. In this case although $T^u_{\eta_p}=U\sqrt{\boldsymbol{\eta_p}}$ takes  a nondiagonal form with  matrix elements $[T^u_{\eta_p}]_{ij}=\sqrt{\eta_{p_j}}\bra{e_i}U\ket{e_j}$, the positive matrix  ${T^{u^{\dagger}}_{\eta_p}}  T^u_{\eta_p}$ is still diagonal and equal to $\boldsymbol{\eta_p}$.
Consequently,     to each fiber $T^e_{\eta_p}\in \mathcal{F}$ over $\boldsymbol{\rho_{_T}}\in \mathcal{M}$   one can  associate the corresponding  fiber $\sqrt{\boldsymbol{\eta_p}}\in \mathcal{F}$ whose projection downs to the  diagonal state $\boldsymbol{\eta_p}\in \mathcal{M}$.
 This particular fiber $\sqrt{\boldsymbol{\eta_p}}\in \mathcal{F}$, and its corresponding diagonal state $\boldsymbol{\eta_p}\in \mathcal{M}$, are special in the sense that they correspond to a perfect discrimination problem, i.e.  the set can be discriminated unambiguously with probability one. In this case the states to be discriminated are mutually orthogonal and can be distinguished by a classical observer.

It is worth to note here  that parameterizing quantum states using a  map similar to Eq.  \eqref{PiMap} was also considered already. Obviously, different parametrization of  $T$ leads to  different parametrization  of  quantum state $\boldsymbol{\rho_{_T}}$.  Very recently, the authors of \cite{IlinJPA2018} defined $T$ as a nonzero Hermitian matrix parameterized by a real $(N^2-1)$-dimensional vector,  and introduced a   parametrization of quantum states.  The essence of our method  is that  the matrix $T$ naturally parameterized in terms of the $N^2-1$ real parameters of the discrimination problem; the $N(N-1)/2$ complex parameters $\gamma_{ij}=\bra{\phi_i}\phi_j\rangle$  provided  by  the mutual overlap of the states to be discriminated,  and the rest  $N-1$ real parameters given by the prior probabilities $\{\eta_i\}_{i=1}^{N}$ (recalling the normalization condition $\sum_{i=1}^N \eta_i=1$).

\subsection{Two-level system}
In order to see how the  procedure work let us consider the   simplest case  $N=2$.
For a given two linearly independent pure states $\ket{\phi_1}$ and $\ket{\phi_2}$ with the  prior probabilities $\eta_1$ and $\eta_2$, respectively,  one can use the familiar Gram-Schmidt procedure \cite{ArfkenBook2013} to construct a specific orthonormal basis $\ket{e_1}$ and $\ket{e_2}$ as
\begin{eqnarray}
\ket{e_1}= \ket{\phi_1},\qquad \ket{e_2}=\frac{ -\gamma_{12}\ket{\phi_1}+\ket{\phi_2}}{\sqrt{1-|\gamma_{12}|^2}},
\end{eqnarray}
where $\gamma_{12}=\bra{\phi_1}\phi_2\rangle$.
Using this  orthonormal basis, we can represent $\ket{\phi_1}$ and $\ket{\phi_2}$ as
\begin{eqnarray}
\ket{\phi_1}= \ket{e_1},\qquad \ket{\phi_2}=\gamma_{12}\ket{e_1}+\sqrt{1-|\gamma_{12}|^2}\ket{e_2}.
\end{eqnarray}
 One can also express  the dual set $\tilde{\Phi}=\{\ket{\tilde{\phi}_1},\ket{\tilde{\phi}_2}\}$ as
\begin{eqnarray}
\ket{\tilde{\phi}_1}=\ket{e_1}-\frac{\gamma_{12}^\ast}{\sqrt{1-|\gamma_{12}|^2}}\ket{e_2},\quad \ket{\tilde{\phi}_2}=\frac{1}{\sqrt{1-|\gamma_{12}|^2}}\ket{e_2}.
\end{eqnarray}
In this orthonormal basis $\{\ket{e_1},\ket{e_2}\}$,  $T^e_{\eta_p}$ is represented as $[T^e_{\eta_p}]_{kl}=\sqrt{\eta_{p_k}}\bra{\tilde{\phi}_k}e_l\rangle$, so that  for a particular permutation $\eta=\diag\{\eta_1,\eta_2\}$, we get
\begin{equation}\label{T222}
T^e_\eta=
\begin{pmatrix}
     \sqrt{\eta_1} & \frac{-\gamma_{12}\sqrt{\eta_1}}{\sqrt{1-|\gamma_{12}|^2}} \\
    0      &  \frac{\sqrt{\eta_2}}{\sqrt{1-|\gamma_{12}|^2}}
\end{pmatrix},
\end{equation}
where can be used to find the following parametrization for an arbitrary density matrix of qubit systems
\begin{equation}\label{Rho-N=2}
\boldsymbol{\rho_{_{T}}}=\left(\begin{array}{cc} \eta_1(1-|\gamma_{12}|^2)& -\gamma_{12} \eta_1\sqrt{1-|\gamma_{12}|^2} \\
  -\gamma_{12}^\ast \eta_1\sqrt{1-|\gamma_{12}|^2}      &  \eta_1|\gamma_{12}|^2+\eta_2
\end{array}\right).
\end{equation}

\subsection{$N$-level system}
To  proceed further with a system of $N$  pure states,   we temporary change our notation and  use $T^{(N)}$ instead of $T_{\eta_p}^e$ as a linear invertible transformation that construct the orthonormal set $\{\ket{e_i}\}$ from a priori nonorthogonal
linearly independent set $\{\ket{\phi_i}\}$, via the Gram-Schmidt procedure
\begin{equation}
T^{(N)}\ket{\phi_i}=\ket{e_i}, \quad \mbox{for} \;\;  i=1,\cdots,N.
\end{equation}
Let $a_{ni}$, for $i=1,\cdots,n$ and $n=1,\cdots,N$, denote the coefficients of the expansion of $\ket{\phi_n}$ in terms of the above orthonormal basis, i.e. $\ket{\phi_{n}}=\sum_{i=1}^{n}a_{ni}\ket{e_i}$,
where $a_{nn}=\sqrt{1-\sum_{i=1}^{n-1}|a_{ni}|^2}$ guaranties the normalization of $\ket{\phi_n}$. We then  have an explicit relation  for $T^{(N)}$ in terms of $T^{(N-1)}$ as
\begin{eqnarray}
T^{(N)}=\left(\begin{array}{c|c}T^{(N-1)} & O \\ \hline \\ O^\T & 1 \end{array} \right)
\left(\begin{array}{c|c} \mathbb{1}_{N-1} & A^{(N-1)} \\ \hline \\ O^\T & 1/{a_{NN}} \end{array} \right).
\end{eqnarray}
Here, $O=(0,\cdots,0)^\T$ denotes an $(N-1)$-components zero vector, $O^\T$ is its transpose, and $\mathbb{1}_{N-1}$ is the unit matrix with dimension $N-1$. Moreover $A^{(N-1)}$ is an $(N-1)$-components vector defined by $A^{(N-1)}=(-a_{N,1},-a_{N,2},\cdots, -a_{N,N-1})^\T/a_{N,N}$.
Recalling that for $N=2$ we have
\begin{equation}
T^{(2)}=\left(\begin{array}{cc}
  1 & \frac{-\gamma_{12}}{\sqrt{1-|\gamma_{12}|^2}} \\
  0 & \frac{1}{\sqrt{1-|\gamma_{12}|^2}}
\end{array}\right),
\end{equation}
the above equation allows us to obtain $T^{(N)}$ for an arbitrary value of $N$. Now, having $T^{(N)}$ for an arbitrary $N$, one can simply obtain  the matrix  $T_{\eta_p}^e$ by  $T_{\eta_p}^e=\sqrt{\boldsymbol{\eta_p}}T^{(N)}$.
For further use  consider the case of three pure states, i.e.  $N=3$,  to get
\begin{equation}\label{T3}
T^{(3)}=
\left(\begin{array}{ccc}
   1     &  \frac{-\gamma_{12}}{\sqrt{1-|\gamma_{12}|^2}} &  \frac{\gamma_{12}\gamma_{23}-\gamma_{13}}{(1-|\gamma_{12}|^2)\sqrt{1-|\gamma_{13}|^2-|a_{23}|^2}}  \\
    0      & \frac{1}{\sqrt{1-|\gamma_{12}|^2}} & \frac{\gamma_{23}-\gamma_{21}\gamma_{13}}{(1-|\gamma_{12}|^2)\sqrt{1-|\gamma_{13}|^2-|a_{23}|^2}} \\
0 & 0 & \frac{1}{\sqrt{1-|\gamma_{13}|^2-|a_{23}|^2}}
\end{array}\right),
\end{equation}
where $a_{23}=(\gamma_{23}-\gamma_{21}\gamma_{13})/\sqrt{1-|\gamma_{12}|^2}$.

\section{Application to quantum state discrimination}\label{section-3}
Consider a  quantum system prepared
in one of the states $\Phi=\{\ket{\phi_i}\}_{i=1}^N$  with a priori probability $\{\eta_i\}_{i=1}^N$.  An  unambiguous discrimination of this set  requires $N+1$ POVM elements $\{E_k\}_{k=0}^{N}$  such that  $\{E_k\}_{k=1}^{N}$ is required for $N$  pure states $\{\ket{\phi_k}\}_{k=1}^{N}$ with no-error condition $\langle \phi_j | E_k | \phi_j \rangle = 0$  ($ k \neq j$), and  $E_0=I-\sum_{k=1}^{N}E_k$ correspond to  the inconclusive result \cite{CheflesPLA1998}.  The  aim is to minimize the probability of inconclusive result, or equivalently, to maximize success probability  $p_{\textrm{success}}=\sum_{k=1}^{N}\eta_k\bra{\phi_k}E_k\ket{\phi_k}$, with optimization taken over all possible POVMs.
The optimization procedure involved in the definition of  $p^{\max}_{\textrm{success}}$ prevents one to write an analytical expression for the maximum probability of success of an arbitrary discrimination problem except for some simple cases such as the case of two pure states with arbitrary probability \cite{JaegerPLA1995} and a symmetric set with equal probabilities \cite{CheflesBarnettPLA1998}. Indeed, it is shown that  the problem of finding an optimal measurement can be formulated as a semidefinite programming \cite{Eldar2004,JafarizadehPRA2008}. In \cite{PangPRA2009}, the authors  have  studied the problem of optimum
unambiguous discrimination of $N$ linearly independent pure states and  have derived some analytical properties of
the optimum solution.
Moreover, for $N>2$ pure states, a   geometric view  for the optimal unambiguous discrimination \cite{BergouPRL2012} and  an upper bound for the optimum  probability of success \cite{BandyopadhyayPRA2014} are also considered.
The problem of finding the set of quantum states that can be deterministically discriminated is analyzed by Markham \etal \cite{MarkhamPRA2008}. From a geometric point of view, they have shown that this problem is equivalent to that of embedding a simplex of points whose distances
are maximal with respect to the Bures distance or trace distance.

The above discussion reveals that despite much efforts devoted to the problem of quantum state discrimination, the subject  yet needs more attention as  calculating  the  maximum probability of success is not an easy task to handle analytically. Accordingly, having a more computable quantity could be beneficial in order to have estimation on  the extent to which a given set can be discriminated.  With this as our  motivation, we invoke the framework of section \ref{section-2} and associated a  measure  of discriminability to each discrimination problem. This quantity, though distinct from the maximum probability of success, can  be useful in estimating the amount of discriminability of a given set  as it captures   information on the problem.

In section  \ref{section-2} we show that  to each discrimination problem $\Phi_\eta$, consisting of $N$ linearly independent  pure quantum states $\Phi=\{\ket{\phi_i}\}$ and  the corresponding  occurrence probabilities $\eta=\{\eta_i\}$, one can associate,  up to a permutation over the probabilities $\{\eta_i\}$,  a unique pair of  density matrices  $\boldsymbol{\rho_{_{T}}}$ and $\boldsymbol{\eta_{{p}}}$   defined on the $N$-dimensional Hilbert space $\mathcal{H}_N$. This  pair of states  is obtained from projection of the pair of operators $T^u_{\eta_p}$ and $\sqrt{\boldsymbol{\eta_p}}$, respectively, following the route of map \eqref{PiMap}. Clearly, if the set is  completely discriminable these operators are equal, up to unitary transformation, however, any deviation from perfect discriminability makes them different in nature.  Motivated by this,  we define the distance between two states $\boldsymbol{\rho_{_T}}$ and $\boldsymbol{\eta_p}$ as the length of the shortest path between two operators $T^u_{\eta_p}$ and $\sqrt{\boldsymbol{\eta_p}}$ \cite{ZyczkowskiBook2006}.    The squared Hilbert-Schmidt distance of $T^u_{\eta_p}$ and $\sqrt{\boldsymbol{\eta_p}}$ gives us
\begin{eqnarray}\nonumber
D^2_B \left(\boldsymbol{\rho_{_T}} , \boldsymbol{\eta_p}\right)
&=&\min_{U} \left\|\frac{UT^e_{\eta_p}}{\sqrt{{\Tr[{T^{e^\dagger}_{\eta_p}}T^e_{\eta_p}]}}}-\sqrt{\boldsymbol{\eta_p}}\right\|_2^2 \\
&=& 2-\sqrt{F(\boldsymbol{\rho_{_T}},\boldsymbol{\eta_p})},
\end{eqnarray}
where
\begin{equation}
F(\boldsymbol{\rho_{_T}},\boldsymbol{\eta_p})=\left[\Tr\sqrt{\sqrt{\boldsymbol{\rho_{_T}}}\boldsymbol{\eta_p} \sqrt{\boldsymbol{\rho_{_T}}}}\right]^2,
\end{equation}
denotes the fidelity between  $\boldsymbol{\rho_{_T}}$ and the diagonal state $\boldsymbol{\eta_p}$. By construction, $D_B(\boldsymbol{\rho_{_T}},\boldsymbol{\eta_p})$ is nothing but the  Bures distance between $\boldsymbol{\rho_{_T}}$ and $\boldsymbol{\eta_p}$ \cite{ZyczkowskiBook2006}.
If the original set $\Phi$  be orthonormal, then $\boldsymbol{\rho_{_T}}=\boldsymbol{\eta_p}$, and we get $F(\boldsymbol{\rho_{_T}},\boldsymbol{\eta_p})=1$. Note that for a given discrimination problem $\Phi_\eta$ both associated density matrices $\boldsymbol{\rho_{_T}}$ and $\boldsymbol{\eta_p}$ are unique up to a permutation over the set of the  prior probabilities $\{\eta_i\}_{i=1}^{N}$. Having this in mind, we   define  the following quantity as a measure of how and in what extent the set $\Phi_{\eta}=\{\ket{\phi_i},\eta_i\}_{i=1}^N$ is discriminable
\begin{equation}
\mathcal{D}(\Phi_{\eta})=\max_{p}F(\boldsymbol{\rho_{_T}},\boldsymbol{\eta_p}),
\end{equation}
where $\max_{p}$ is  taken  over all permutations    $p=\{p_1, \cdots,p_N\}$  of $\{1,\cdots,N\}$, recalling  that $\boldsymbol{\rho_{_T}}$ is also depend on the permutation $p$.   For a given set $\{\ket{\phi_{i}}\}$, the discriminability $\mathcal{D}(\Phi_{\eta})$ reaches  its minimum value if and only if $\eta_i=\frac{1}{N}$ for $i=1,\cdots,N$. In this case we have
\begin{equation}
\mathcal{D}(\Phi_{\eta=\frac{1}{N}})=\min_{\eta}\mathcal{D}(\Phi_{\eta})=\frac{1}{N}\left[\Tr\sqrt{\boldsymbol{\rho_{_T}}}\right]^2.
\end{equation}
Using this and the fact that $\Tr\sqrt{\boldsymbol{\rho_{_T}}}\ge \Tr{\boldsymbol{\rho_{_T}}}\ge 1$, we find that $\frac{1}{N} \le \mathcal{D}(\Phi_{\eta}) \le 1$.  The lower bound $\mathcal{D}(\Phi_{\eta})=\frac{1}{N}$  is attained  if and only if  the set becomes dependent. On the other hand the full discriminability, i.e.  $\mathcal{D}(\Phi_{\eta})=1$,    happens if and only if the  set  $\Phi$ be an orthonormal set, i.e. if and only if the set can be discriminated unambiguously with probability one.
Furthermore $\mathcal{D}(\Phi_{\eta})$ is invariant under any unitary transformation $V$ performed on the set $\Phi$, i.e. $\mathcal{D}(\Phi^V_{\eta})=\mathcal{D}(\Phi_{\eta})$ for $\Phi^V=\{V\ket{\phi_i}\}$. To see this note that the associated density matrix $\boldsymbol{\rho_{_T}}$ is expressed in terms of the overlap parameters $\gamma_{ij}=\bra{\phi_i}\phi_j\rangle$ and the probabilities $\{\eta_i\}$,  both are invariant under such transformations. Another way to see this is the use of the fact the associated pair of density matrices $\boldsymbol{\rho_{_{T}}}$ and $\boldsymbol{\eta_p}$ transform as $\boldsymbol{\rho_{_{T}}} \rightarrow V\boldsymbol{\rho_{_{T}}}V^\dagger$ and $\boldsymbol{\eta_p} \rightarrow V\boldsymbol{\eta_p} V^\dagger$ under such transformation, but this  leaves the fidelity between them invariant. In the following we provide some examples and compare the maximum success probability $P^{\max}_{\textrm{success}}$ with the normalized discriminability, i.e. $(N\mathcal{D}(\Phi_{\eta})-1)/(N-1)$.

\subsection{Two pure states}
\begin{figure}
\mbox{\includegraphics[scale=0.4]{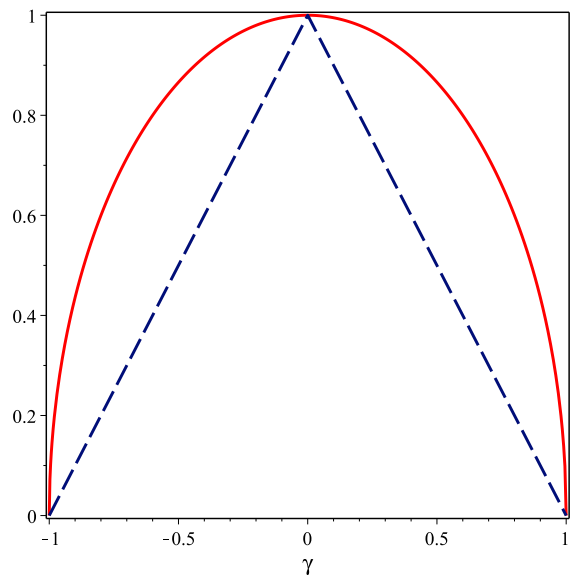}}
\caption{(Color online) Plots of $(2 D(\Phi_{\eta})-1)$ (solid-red line) and $P_{\textrm{IDP}}$ (dashed-blue line) in terms of $\gamma_{12}=\bra{\phi_1}\phi_2\rangle$ (for real $\gamma_{12}$).}
\label{fig1}
\end{figure}

\begin{figure}
\mbox{\includegraphics[scale=0.4]{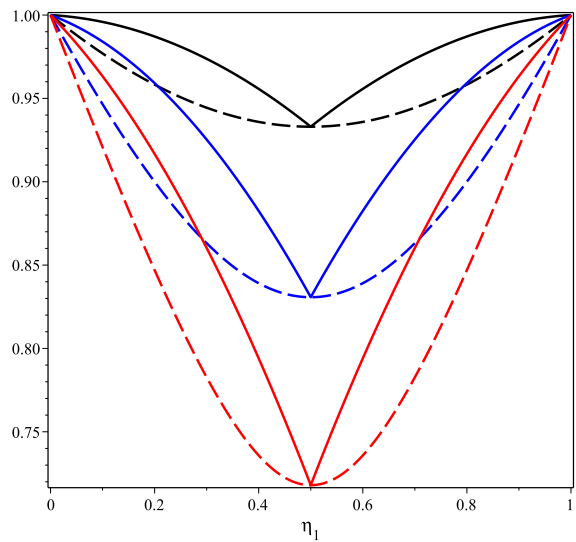}}
\caption{(Color online) Graphs showing $\mathcal{D}(\Phi_{\eta})$ (solid lines) and $P_{\textrm{correct}}$ (dashed lines) as a function of $\eta_1$ for  $\gamma_{12}=\frac{1}{2}$ (black lines), $\gamma_{12}=\frac{3}{4}$ (blue lines), and $\gamma_{12}=\frac{9}{10}$ (red lines). }
\label{fig2}
\end{figure}

For the case with two pure states, $\boldsymbol{\rho_{_{T}}}$ is given by Eq. \eqref{Rho-N=2} and  $\boldsymbol{\eta_p}=\diag\{\eta_1,\eta_2\}$.
In this case using $F(\rho_1,\rho_2)=\Tr{\rho_1\rho_2}+2\sqrt{\det{\rho_1}\det{\rho_2}}$ \cite{ZyczkowskiBook2006}, we get $F(\boldsymbol{\rho_{_{T}}},\boldsymbol{\eta_p})=\eta_1^2+\eta_2^2+\eta_1|\gamma_{12}|^2(\eta_2-\eta_1)+2\eta_1\eta_2\sqrt{1-|\gamma_{12}|^2}$.  We therefor arrive at the following relation for the discriminability of two pure states
\begin{eqnarray}\label{D-TwoPure}
\mathcal{D}(\Phi_{\eta})=\eta_1^2+\eta_2^2+2\eta_1\eta_2\sqrt{1-|\gamma_{12}|^2}+|\gamma_{12}|^2(\eta_1\eta_2-\eta_{\min}^2),
\end{eqnarray}
where $\eta_{\min}=\min\{\eta_1,\eta_2\}$.
For the case with equal prior probabilities $ \eta_1=\eta_2=\frac{1}{2} $, this quantity becomes
\begin{equation}
\mathcal{D}(\Phi_{\eta=\frac{1}{2}})=\frac{1}{2} (1+\sqrt{1-|\gamma_{12}|^2}),
\end{equation}
which is, surprisingly,   equal to the probability of correction in minimum-error discrimination \cite{HelstromBook1976}. Figure \ref{fig1} compares the plot of $\left(2\mathcal{D}(\Phi_{\eta})-1\right)$ with the maximum probability of success given by $p_{\textrm{success}}^{\max}=1-|\bra{\phi_1}\phi_2\rangle|$ \cite{IvanovicPLA1987,PeresPLA1988,DieksPLA1988}.
For general case with unequal probabilities, we  compare discriminability $\mathcal{D}(\Phi_{\eta})$ with probability of correction in minimum-error   discrimination. As a result, two quantities reach   their common lowest value at the same point $ \eta=\frac{1}{2} $. Moreover, discriminability has been decreased by increasing the overlap parameter $\gamma_{12}$ (Fig.\ref{fig2}).

\subsection{Three Pure States}
\begin{figure}
\mbox{\includegraphics[scale=0.4]{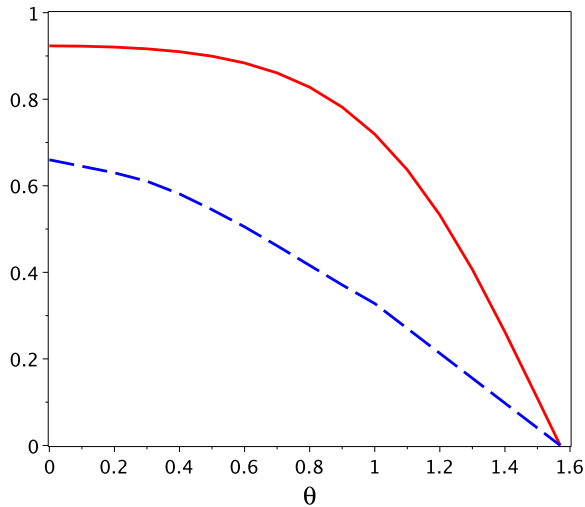}}
\caption{(Color Online) Plots show  ($ 3D(\eta,\Phi)-1) /2$ (solid-red line) and probability of success (dashed-blue line).}
\label{Fig3}
\end{figure}

For the case of three pure states, we  can use Eq. \eqref{T3} and write an expression for density matrix $\boldsymbol{\rho_{_{T}}}$.  We  continue  with the particular case that all $\gamma_{ij}$ are real, so that $\gamma_{12}=\cos{\alpha}$, $\gamma_{13}=\cos{\phi}\sin{\theta}$, and $\gamma_{23}=\sin{\theta}\cos{(\alpha-\phi)}$. By setting  $\alpha=\pi/3$ and $\phi=\pi/4$, we can  provide a  comparison between the discriminability and the  maximum probability of success   \cite{SugimotoPRA2010} for the case with equal prior probabilities.  This comparison shows a great consistency between two quantities (Fig. \ref{Fig3}).

\section{CONCLUSION}
By starting from  a given set of linearly independent nonorthogonal states $\{\ket{\phi_i}\}$, with the corresponding  prior  probabilities $\{\eta_i\}$, we could obtain a set of orthogonal states $\{\ket{e_i}\}$. To this aim, we define the  invertible transformation matrix  $T^e_{\eta_p}$ which is not unique in the sense that it  defines the orthonormal basis $\{\ket{e_i}\}$ up to a unitary transformation. Moreover, it is also far from uniqueness to the extent of a permutation of  the prior probabilities $\{\eta_i\}$. However, this transformation allows us to associate to each nonorthogonal set $\{\ket{\phi_i}\}$ and the corresponding  prior  probabilities $\{\eta_i\}$  a unique pair of  density matrices  $\boldsymbol{\rho_{_{T}}}$ and $\boldsymbol{\eta_{{p}}}$, up to a permutation over the probabilities $\{\eta_i\}$. The first one, $\boldsymbol{\rho_{_{T}}}$, is of particular importance as it provides a new parametrization for a general density matrix supported on $\mathcal{H}_N$.  Interestingly, it captures all the information about the quantum state discrimination problem, i.e. one can extract all the mutual overlaps  $\gamma_{ij}=\bra{\phi_i}\phi_j\rangle$ and  the occurrence probabilities $\{\eta_i\}$ from it. The second one, $\boldsymbol{\eta_{{p}}}$, on the other hand, represents a  diagonal density matrix and captures the information about a perfect quantum state  discrimination problem, i.e.   when the states to be discriminated are mutually orthogonal and can be distinguished by a classical observer.  Using this pair of density matrices and the notion of fidelity, $F(\boldsymbol{\rho_{_{T}}}, \boldsymbol{\eta_{{p}}})$, we  define  a quantity called discriminability which shows the possibility of discrimination among a set of quantum states. This quantity is as easy to calculate as fidelity and can find useful applications in estimating the  extent to which a given set of pure states can be discriminated. The existence of nonorthogonal states could be responsible for some nonclassicalities  such as quantum discord, no-cloning theorem and  no-local-broadcasting which have widespread applications in quantum information theory.  The approach presented in this paper associates a distinguishability problem to each discrimination problem. We hope  the method presented in this work  should shed some light on the similar  situations which we deal with quantum properties arising from non orthogonality.

\acknowledgments
This work was supported by Ferdowsi University of Mashhad under Grant No.  3/42098 (1395/08/08).

\end{document}